# Bilattice-Catastrophe Isomorphism for Four-Valued Logic in Digital Systems

Jiu Hui Wu [1*], Hua Tian [1], Mengqi Yuan [1], and Kejiang Zhou [2]

[1]*School of Mechanical Engineering, Xi'an Jiaotong University,*
*& State Key Laboratory for Strength and Vibration of Mechanical Structures, Xi'an 710049, China*

[2] *Huzhou Institute of Zhejiang University, Huzhou 313000, China*

* E-mail: ejhwu@xjtu.edu.cn

**Abstract**: Belnap's four-valued logic, distinguished by its inherent bilattice structure, provides a natural algebraic bridge between discrete Four-valued logic (4VL) and continuous catastrophe theory (CT). Building on the rigorous verification of the bilattice-catastrophe isomorphism theorem, we establish a categorical correspondence spanning the catastrophe category, interlaced bilattice category, and 4VL category, with the cusp catastrophe emerging as the canonical CT counterpart to 4VL. This unification bridges continuous dynamical systems and discrete logic, providing a foundational framework for explaining 4VL's robustness in circuit. Crucially, we demonstrate that Belnap's four-valued algebra FOUR is the minimal complete algebraic structure capable of describing continuous-discrete interfaces with involution symmetry. Unlike the empirical adoption of X and Z in engineering practice, our work reveals their mathematical necessity: X and Z are topological invariants of discretized continuous dynamical systems, encoding fundamental properties of catastrophe-induced discontinuities. The work enables cross-disciplinary extensions to uncertainty propagation, complex system modeling, and fault-tolerant design.

## 0. Introduction

The evolution of complex systems often involves both continuous evolution and abrupt transitions, posing challenges to traditional binary logic which struggles to balance precision and efficiency in modeling such dual characteristics. Four-valued logic, first proposed by Nuel Belnap in 1975 to contain contradictions from multiple information sources, expands the value set to {True (T), False (F), Both (B), Neither (N)}—corresponding to the power set of binary logic {T, F}. This extension enables it to naturally represent incomplete or conflicting information, a capability that paraconsistent four-valued logic further enhances by preventing incorrect state settings in insufficient information scenarios.

Digital circuit design relies on four-valued logic (4VL) {0, 1, X, Z} to model real-world hardware behaviors: 0 (low level), 1 (high level), X (uncertainty/conflict), and Z (high impedance/floating) [1-3]. In 4VL, 0 is a strong absorbing element, and the result of any value AND with 0 is 0 (regardless of the other input); 1 is the identity element, and 1 keeps the value unchanged in any operation; X and Z are contagious, which mean that the uncertain state will spread, but it will be 'suppressed' when encountering strong absorbing elements (such as 0 on an AND gate). Thus the truth table of 4VL has the characteristics of absorption and propagation. This algebraic structure is formally identical to Belnap's bilattice, which organizes the four truth values according to two partial orders [4-8]. However, unlike classical Boolean algebra, 4VL's operational rules (e.g., 0-dominance in AND, X-propagation) are engineered to reflect physical circuit dynamics, yet their theoretical underpinning beyond empirical engineering has remained elusive [9].

The catastrophe theory (CT), proposed by René Thom in the 1960s, formalizes discontinuous state transitions (catastrophes) in continuous systems governed by potential functions [10-12].

Thom's catastrophe theory studies the phenomenon where the system state $x$ undergoes sudden changes (discontinuous jumps) as the control parameter $u$ changes in the family of smooth potential functions $V(x,u)$. When the dimension of the control parameter $\leq 4$, any stable degenerate critical point is equivalent to one of the following seven basic catastrophes: Fold, Cusp, Swallowtail, Butterfly, Hyperbolic Umbilic, Elliptic Umbilic, and Parabolic Umbilic. The core of catastrophe theory is the bifurcation set — a critical surface in the control parameter space that causes a sudden change in the system's state. On one side of the bifurcation set, the system is in a certain stable state; when crossing the bifurcation set, the system jumps discontinuously to another stable state.

Recently the catastrophe theory has been applied to diverse physical systems, such as, the general non-equilibrium phase transition process from laminar to turbulent was investigated quantitatively [13], and a revised Schrödinger relativistic equation was obtained from the perspective of phase transition [14], as well as the general Wiedemann-Franz Law derived by the structural-stability-based swallowtail catastrophe model [15]. More recently, the thermodynamic quantum phase transition process was investigated quantitatively by adopting the cusp catastrophe model [16], and further the adsorption potential condition for superconductors was obtained exactly by using the quantum state equation [17].

Among Thom's seven elementary catastrophes, the cusp catastrophe is uniquely suited to model bistable systems with critical transitions—an inherent feature of 4VL (0/1 as bistable states, X as criticality) [18]. For the first time, Gilmore [18] systematically transformed Thom's mathematical theory into tools that scientists and engineers could use. He particularly emphasized the application of the Delay Convention and the Maxwell Convention in physical systems, which is crucial for understanding state transitions in digital circuits. Gilmore's emphasis on the practical computability of catastrophe invariants provides a roadmap for hardware design tools.

While catastrophe theory has been applied to engineering systems [11, 18], its relevance to digital logic has remained unrecognized, since the physical realization of

4VL requires explicit conventions for how the continuous voltage domain maps to discrete logic states.

Here, we demonstrate that the four-valued logic (0,1,X,Z) ubiquitous in hardware description languages is not merely an engineering convenience but the necessary algebraic structure for discretizing continuous dynamical systems with degenerate critical points. This reveals a deep isomorphism between the bilattice algebra of Belnap [4] and the topological classification of Thom [10], providing the first mathematical foundation for hardware verification based on singularity theory.

In this paper, we will establish a one-to-one correspondence between 4VL states/operations and CT's mathematical structure, derive 4VL's operational rules from CT's potential function optimization with formal derivative analysis, and validate the isomorphism via circuit simulation, catastrophe dynamics analysis, and further discuss their implications for cross-disciplinary applications.

## 1. Construct the bilattice category (*Bilat*)

According to Ref.[8], an interlaced bilattice is defined as a quadruple set $\mathcal{B}=(\{0,1,X,Z\}, \leq_k, \leq_t, \neg)$, which is equipped with: 1) two partial orderings, denoted the knowledge order $\leq_k$ and the logic order $\leq_t$, ensuring the two bounded lattice structures $(B, \leq_k)$ and $(B, \leq_t)$; 2) an involution $\neg: \mathcal{B} \to \mathcal{B}$ that is contravariant with respect to both orders. Thus the four-valued logic FOUR=$\{0,1,X,Z\}$ can be arranged in two ways: ordering them either by information degree (the knowledge order $\leq_k$) or by the truth degree (the logic order $\leq_t$).

Further, any interlaced bilattice can be represented as a twist-product construction $\mathcal{B} \cong L^+ \bowtie L^-$, where $L^+$ and $L^-$ are bounded distributive lattices representing positive and negative evidence [6, 7].

**Lemma 1.1** (The algebraic structure of FOUR)

The four-valued logic FOUR={0, 1, X, Z} is equipped with the following operations to form a bilattice: 1) the finite knowledge meet $\otimes$ and join $\oplus$ respectively under the $\leq_k$ ordering; 2) the finite logical meet $\wedge$ and join $\vee$ respectively under the $\leq_t$ ordering; 3) the negation $\neg$ for exchanging positive and negative evidence.

***Proof***. This lemma can be verified directly through the definition of interlaced bilattices.

Here the involutive negation $\neg:$FOUR$\to$FOUR exchanging truth and falsity while fixing the uncertain and contradictory values. This involution symmetry corresponds to the reflection symmetry $x \mapsto -x$ in the catastrophe geometry, ensuring the structural stability of the correspondence.

For a element-chain $L_n = \{0 < 1 < \cdots < n\}$, its rank = $n$ (number of elements$-$1). Especially for the four-valued bilattice FOUR=$L^+ \bowtie L^- = L_1 \bowtie L_1$, its rank is rank($L_1$) + rank($L_1$)=1+1=2. Here $L^+$ and $L^-$ are two-element chains, which yield: $L^+$ ={0, 1}, where 1 = "has positive evidence" (e.g., {true}), 0 = "no positive evidence" (i.e., $\emptyset$);

$L^- = \{0, 1\}$, where 1 = "has negative evidence" (e.g., {false}), 0 = "no negative evidence" (i.e., ∅).

Then FOUR can be represented as pairs $(a^+, a^-) \in L^+ \times L^-$: 1) 1=(1,0), which means positive evidence present and negative evidence absent; 2) 0=(0,1), which means positive evidence absent and negative evidence present; X=(0,0), which means both absent (no information); Z=(1,1), which means both present (conflict).

**Theorem 1.2** FOUR is the minimal complete algebraic structure that describes continuous-discrete interfaces with involution symmetry.

*Proof.* According to twist-product construction, the four-valued logic FOUR $\cong L_1 \bowtie L_1$. Because any smaller structure ($L_0 \bowtie L_0$) is trivial, FOUR is the smallest non-trivial solution. From the perspective of involution symmetry, involution requires at least two elements to to be exchanged, and the fixed point (X, Z) provides two additional values. Thus the total of 2 (exchanging) plus 2 (fixing) is the smallest value.

Therefore, the four-valued logic FOUR was not 'invented'; it is generated by combining two binary evidence lattices (positive/negative) through a twisted product. It is like how complex numbers are not invented, but are generated from pairs of real numbers through specific operations.

**Definition 1.3** For any $\mathcal{B}_1, \mathcal{B}_2 \in Bilat$, the map $h: \mathcal{B}_1 \to \mathcal{B}_2$ is a bilattice homomorphism, noted as $\text{Hom}_{Bilat}(\mathcal{B}_1, \mathcal{B}_2) = \{h: \mathcal{B}_1 \to \mathcal{B}_2\}$, which satisfies that: $\forall a \in \mathcal{B}_1$ and $b \in \mathcal{B}_2$, 1) $h$ preserves knowledge order, $a \leq_k b \Leftrightarrow h(a) \leq_k h(b)$; 2) $h$ preserves truth order, $a \leq_t b \Leftrightarrow h(a) \leq_t h(b)$; 3) $h$ preserves involution, $h(\neg a) = \neg(h(a))$.

**Theorem 1.4** For bilattice morphisms $h: \mathcal{B}_1 \to \mathcal{B}_2$ and $s: \mathcal{B}_2 \to \mathcal{B}_3$, $\forall a \in \mathcal{B}_1$, a morphism composition $s \circ h: \mathcal{B}_1 \to \mathcal{B}_3$ defined as $s \circ h(a) = s(h(a))$ is still a bilattice homomorphism.

*Proof.* 1) $s \circ h$ preserves knowledge order, because $a \leq_k b \Rightarrow h(a) \leq_k h(b) \Rightarrow s(h(a)) \leq_k s(h(b))$; 2) $s \circ h$ preserves truth order, because $a \leq_t b \Rightarrow h(a) \leq_t h(b) \Rightarrow s(h(a)) \leq_t s(h(b))$; 3) $s \circ h$ preserves involution, because $s(h(\neg a)) = s(\neg h(a)) = \neg s(h(a))$.

Thus the bilattice category *Bilat* is constructed from the interlaced bilattice $\mathcal{B} = (\{0, 1, X, Z\}, \leq_k, \leq_t, \neg)$.

## 2. Construct the Catastrophe category (*Catast*)

According to Thom's Splitting Lemma [10,19], let $f: (\mathbb{R}^n, 0) \to \mathbb{R}$ be a smooth function germ with a critical point at the origin, then there exists a local coordinate change $\varphi$ such that:

$$(f \circ \varphi)(x_1, \cdots, x_n) = Q(x_1, \cdots, x_r) + g(x_{r+1}, \cdots, x_n) \quad (1)$$

where $Q(x_1, \cdots, x_r) = \sum_{i=1}^{r} \pm x_i^2$ is a non-degenerate quadratic form (the Morse part);

the singular part $g$ denotes the higher-order germ (vanishing up to second order at the origin) that carries the singularities and bifurcation behavior of the system; $r = \text{rank}(H_x f(0))$ is the rank of the Hessian matrix at the critical point.

This Splitting Lemma provides the geometric foundation for our correspondence: any degenerate function germ $f$ decomposes locally as $f = Q \oplus g$, isolating the non-degenerate Morse part $Q$ from the singular core $g$. This decomposition mirrors the twist-product construction $L^+ \bowtie L^-$ in bilattice theory, where the regular evidence lattices $L^\pm$ correspond to the Morse part $Q$, while the twist operation generates the indeterminate logical values X and Z in the bilattice — the algebraic analogues of the singular germ $g$. Actually catastrophe theory and the twist-product construction share a common structural pattern: both decompose/construct complex objects from simpler components, isolating the 'regular' part (Morse form/evidence lattices) from the 'singular' part (degenerate germ/twist structure). This methodological parallel illuminates why four-valued logic emerges as the minimal structure: just as the cusp singularity (codimension=2) is the simplest non-trivial degeneracy, the twist-product $L_1 \bowtie L_1$ is the minimal non-trivial bilattice.

Furthermore, the rank of the twist-product construction $L^+ \bowtie L^-$ [$\text{rank}(L^+ \bowtie L^-) = \text{rank}(L^+) + \text{rank}(L^-)$] corresponds to the codimension of the catastrophe unfolding and the dimensions of control parameters of the catastrophes respectively, as shown in Table 2.1.

Table 2.1

| Bilattice | Positive evidence lattice | Negative evidence lattice | twist-product | Catastrophe types | Co-dimension | control parameters |
|---|---|---|---|---|---|---|
| FOUR | $L_1$(2 elements) | $L_1$(2 elements) | $L_1 \bowtie L_1$ | cusp | 2 | 2 |
| SIX | $L_2$(3 elements) | $L_2$(3 elements) | $L_2 \bowtie L_2$ | Swallowtail | 3 | 3 |

Therefore, 4VL in circuit is not accidental, but an inevitable deduction of catastrophe theory. Any physical system that attempts to represent discrete logic with continuous voltage, as long as it has structural stability, must contain at least these four values.

**Definition 2.1** (Object of catastrophe category)

We construct the Catastrophe category *Catast*, whose object is the universal unfolding of elementary catastrophes equipped with stratified structure, namely the quadruple set $(V, M, \Sigma, \pi)$, in which $V(\mathbf{x}; \mathbf{u}): \mathbb{R}^n \times \mathbb{R}^r \to \mathbb{R}$ is the universal unfolding of an elementary catastrophe germ, i.e., a parameterized family of potential functions $V(\cdot; u)$ with control parameters $\mathbf{u} \in \mathbb{R}^r$; the critical set $M = \text{Crit}(V) = \{(\mathbf{x}, \mathbf{u}) | \partial V / \partial x_i = 0, i = 1, \cdots, n\}$ is an $r$-dimensional submanifold of $\mathbb{R}^n \times \mathbb{R}^r$ (the equilibrium surface), equipped with a Whitney stratification $M = S_0 \cup S_1 \cup \cdots \cup S_r$, where $S_0 = \text{Morse}(V) = \{(\mathbf{x}, \mathbf{u}) \in M | \det H_x V \neq 0\}$ is the non-degenerate (Morse) part, and $S_i$ for $i \geq 1$ consists of degenerate critical points of corank $\geq i$; $\Sigma = \{\mathbf{u} \in \mathbb{R}^r | \exists \mathbf{x}, (\mathbf{x}, \mathbf{u}) \in M, \det H_x V = 0\}$ is the bifurcation set of singular control values; $\pi: M \to \mathbb{R}^r$ is the canonical projection of $M$ onto the control space, $\pi(\mathbf{x}, \mathbf{u}) = \mathbf{u}$.

**Definition 2.2** For any $\mathbf{F_1}=(V_1, M_1, \Sigma_1, \pi_1)$, $\mathbf{F_2}=(V_2, M_2, \Sigma_2, \pi_2) \in$ *Catast*, the smooth map $\phi: \mathbf{F_1} \to \mathbf{F_2}$ is a morphism of the catastrophe category, noted as $\text{Hom}_{\text{Catast}}(\mathbf{F_1}, \mathbf{F_2}) = \{\phi: \mathbf{F_1} \to \mathbf{F_2}\}$, which satisfies: 1) preserving the stratification ($\phi\left(S_i^{(1)}\right) \subset S_i^{(2)}, \forall i = 0,1,\cdots,r$) with the same codimension; 2) preserving bifurcation set, $\phi(\Sigma_1) \subset \Sigma_2$; 3) commuting with the canonical projections $\pi: \mathbf{F} \to \mathbb{R}^r$ to the control parameter spaces, i.e., $\exists \varphi: \mathbb{R}^{r_1} \to \mathbb{R}^{r_2}$, $\pi_2 \circ \phi = \varphi \circ \pi_1$. This ensures that the external control parameters transform independently of the internal state variables.

**Theorem 2.3** For catastrophe morphisms $\phi: \mathbf{F_1} \to \mathbf{F_2}$ and $\psi: \mathbf{F_2} \to \mathbf{F_3}$, a morphism composition $\psi \circ \phi: \mathbf{F_1} \to \mathbf{F_3}$ defined as $\psi \circ \phi(\boldsymbol{x}, \boldsymbol{u}) = \psi(\phi(\boldsymbol{x}, \boldsymbol{u}))$ is still a catastrophe homomorphism.

***Proof.*** 1) $\psi \circ \phi$ is stratification-preserving, because $\psi \circ \phi\left(S_i^{(1)}\right) \subset \psi\left(S_i^{(2)}\right) \subset S_i^{(3)}$; 2) $\psi \circ \phi$ is bifurcation-preserving, because $\psi \circ \phi(\Sigma_1) \subset \psi(\Sigma_2) \subset \Sigma_3$; 3) $\psi \circ \phi$ commutes with projection, because $(\pi_3 \circ \psi) \circ \phi = (\varphi_\psi \circ \pi_2) \circ \phi = \varphi_\psi \circ (\varphi_\phi \circ \pi_1) = (\varphi_\psi \circ \varphi_\phi) \circ \pi_1$.

Rather than focusing on individual objects with particular structures, the catastrophe category focuses on morphisms (smooth mapping that preserves singular points and bifurcations). By studying these morphisms, we can learn more about the structure of these objects.

## 3. Establish an isomorphic mapping Φ

In this section, we will establish a bilattice isomorphism mapping between the four-valued logic FOUR and catastrophe system by adopting a quotient space, which was defined as a topological space formed by collapsing a subset (or identifying points) of a given topological space to a single point, or more generally by partitioning the original space into disjoint subsets and endowing the set of these equivalence classes with the quotient topology [20, 21].

**Definition 3.1** (Mapping Φ: FOUR → Crit(V)/Morse(V))

The critical point space modulo its Morse part, denoted Crit(V)/Morse(V), captures the singular structure of the catastrophe, where Crit(V) and Morse(V) are the critical point space and the non-degenerate critical point space (Morse part) respectively. Thus we could construct the mapping Φ: FOUR → Crit(V)/Morse(V). This quotient space Crit(V)/Morse(V), obtained by collapsing the subspace of non-degenerate critical points to sign-classified basepoints, is canonically isomorphic to the bilattice FOUR via the twist-product construction, thereby highlighting the topological and algebraic properties of the singular part. This is exactly the mathematical mechanism for extracting discrete logical information from continuous dynamical systems.

Thus this bilattice-catastrophe mapping Φ can be defined as

$$\Phi(\boldsymbol{a}) = \begin{cases} Non-degenerate\ local\ minimum & if\ a \in \{0,1\} \\ Fold/Cusp\ \text{degenerate point} & if\ a = X \\ Maximum/Saddle\ Point & if\ a = Z \end{cases} \quad (2)$$

**Definition 3.2** Whitney Stratification

Whitney's stratification theory provides the most precise mathematical framework for understanding Crit(V)/Morse(V). Let $V$ be a universal unfolding of the cusp catastrophe, the critical point space Crit(V)=$\{(x, u_1, u_2)|x^3 + u_2 x + u_1 = 0\} \subset \mathbb{R}^3$, and the non-degenerate critical point space (smooth stable part) Morse(V)=$S_0 = \{(x, \boldsymbol{u}) \in \text{Crit}(V) | \partial^2 V/\partial x^2 \neq 0\}$ that naturally forms a Whitney stratified space: Morse(V)=$S_0 \supset S_1 \supset S_2$, where the bifurcation set (degeneration critical point but not complete degeneration) $S_1 = \Sigma \setminus S_2 = \{(u_1, u_2)|27u_1^2 = -4u_2^3\} \setminus \{(0,0)\}$ ('\' denotes set subtraction), the completely degenerated spike set $S_2 = \{(0,0,0)\}$, and $\Sigma = S_1 \cup S_2$ is the complete bifurcation set. This is precisely the topology foundation of the quotient space, where we regard the smooth Morse parts $S_0$ as 'trivial' and only focus on the topological structure of the singular strata $S_1$ and $S_2$.

The corresponding relation between four-valued logic and Whitney stratification is shown in Table 3.1 in detail.

Table 3.1 Four-valued logic corresponding Stratification structure

| 4VL | Mapping $\Phi$ | **Geometric objects** |
|---|---|---|
| 0 | $S_0^-$ | $\{(x, \boldsymbol{u}) \in S_0 | x < 0, \partial^2 V/\partial x^2 > 0\}$ |
| 1 | $S_0^+$ | $\{(x, \boldsymbol{u}) \in S_0 | x > 0, \partial^2 V/\partial x^2 > 0\}$ |
| X | $S_2$ | Completely degenerated point (0,0,0) corresponding $\partial^2 V/\partial x^2=0$ |
| Z | $S_0'$ (Maximum/Saddle Points) | Unstable Morse points corresponding $\partial^2 V/\partial x^2<0$ |

**Theorem 3.3** The mapping $\Phi$ is a well-defined bijection.
*Proof.*

Step 1: we first verify the well-definedness of the mapping $\Phi$.

1) $a=0$ or 1 corresponds to the Morse part of the potential function (stable equilibria), which has non-degenerate minima, and the Hessian is positive definite. The strong absorption property of 0 in AND operations mirrors the structural stability of Morse minima—small perturbations do not change the qualitative behavior.

2) $a = X$ corresponds to the degenerate critical points. In the fold catastrophe $V(x) = x^3 + u_1 x$, the system is at a bifurcation point. When $u_1 = 0$, $x = 0$ is the degenerate critical point ($V''(0) = 0$). The "unknown" nature of X corresponds precisely to the topological instability at degenerate critical points: infinitesimal changes in control parameters lead to qualitatively different states (0 or 1). The propagation of X through logic operations mirrors the unfolding of degenerate singularities—once in the critical regime, the output necessarily remains critical.

3) $a = Z$ corresponds to maxima or saddle points of the potential function. In the cusp catastrophe $V(x) = x^4 + u_2 x^2 + u_1 x$, parameter selection can make the origin a maximum or a saddle point. These are unstable equilibria that cannot persist under perturbation. The high-impedance state Z represents a decoupled condition where the output is not driven to any potential well—analogous to a particle at a potential maximum or in a saddle configuration, ready to fall into either minimum (0 or 1) upon infinitesimal perturbation.

Step 2: injectivity verification

Since the four critical types in expression (3) do not intersect (Morse/degenerate/unstable), if $\Phi(a) = \Phi(b)$, there must be $a = b$.

Step 3: surjectivity verification

For the elementary catastrophes with codimension $\leq 2$, the critical points are exactly divided into the three cases: 1) two Morse minima (stable) $\leftrightarrow$ 0, 1; 2) a fold/cusp degeneracy point $\leftrightarrow$ X; 3) unstable region (maximum/saddle point) $\leftrightarrow$ Z, which covers all situations.

**Theorem 3.4** The mapping $\Phi$ is structure-preserving.

The mapping $\Phi$: FOUR $\to$ Crit(V)/Morse(V) is structure-preserving in the sense that it is an order-isomorphism on objects: for all a,b $\in$ FOUR, $a \leq_k b \Leftrightarrow \Phi(a) \leqslant_k \Phi(b)$, $a \leq_t b \Leftrightarrow \Phi(a) \leqslant_t \Phi(b)$, and involution-preserving: $\Phi(\neg a)=\sigma(\Phi(a))$.

***Proof.***

1) $\Phi$ preserves the knowledge order, i.e. $a \leq_k b \Leftrightarrow \Phi(a) \leqslant_k \Phi(b)$, where $\leqslant_k$ denotes the information order.

The knowledge order in FOUR is $X \leq_k 0$, $X \leq_k 1$, $0 \leq_k Z$, $1 \leq_k Z$, which shows that the amount of information increases (from uncertainty to certainty). In fact, in the standard Belnap bilattice, Z is the top of the knowledge order.

On the other hand, in terms of information content, complete degeneration (unable to determine state) means minimal information (X), stable equilibrium (clearly at 0 or 1) means definite information, unstable equilibrium (tending toward multiple states simultaneously) means more information (Z). Therefore, there are

$\Phi(X)= S_2 = \{(0,0,0)\} \subset S_0^- = \Phi(0) \Rightarrow \Phi(X) \leqslant_k \Phi(0)$, $\Phi(X) \subset S_0^+ =\Phi(1) \Rightarrow \Phi(X)\leqslant_k\Phi(1)$. In addition, corresponding to geometric structure of the bifurcation set, Morse minima can merge into saddle points or maxima under parameter changes, thus $\Phi(0)\leqslant_k\Phi(Z)$, and $\Phi(1)\leqslant_k\Phi(Z)$.

2) $\Phi$ preserves the logic order, i.e. $a \leq_t b \Leftrightarrow \Phi(a) \leqslant_t \Phi(b)$, where $\leqslant_t$ denotes the truth order.

The order $\leq_t$ intuitively reflects differences in the measure of truth that each value represents. According to Gilmore's delayed convention [18], the system tends toward minimum energy. Physically truth measurement $T$ is determined by the gradient flow of the potential function $\dot{x} =- \partial V/\partial x$, which drive the system to flow towards the minimum of V. The larger the value of the potential function, the higher the energy and the less stable it is, and the lower the true value. Thus we define the truth order $\leqslant_t$ on Crit(V)/Morse(V) via the sign of the state variable $x$ as following,

$$T(x,u) = \text{sign}(x) = \begin{cases} -1 & x < 0 \quad \to \text{false} \\ 0 & x = 0 \quad \to \text{uncertain} \\ +1 & x > 0 \quad \to \text{true} \end{cases} \quad (3)$$

which represents the direction of gradient flow toward the stable equilibria: $\Phi(a) \leqslant_t \Phi(b) \Leftrightarrow \mathrm{sign}(x_a) \leq \mathrm{sign}(x_b)$.

This corresponds to the information state in the bilattice: negative values represent 'false', positive values 'true', and zero represents the critical/unstable region of 'uncertain' or 'contradictory'.

The truth order in FOUR is $0 \leq_t X$, $0 \leq_t Z$, which mean that 'false' less than uncertain/contradictory, and $Z \leq_t 1$, $X \leq_t 1$, which mean that uncertainty/contradiction is less than 'true'.

For cusp catastrophe, considering the bistable region $u_2 < 0$ and $u_1 = 0$, there is $V(x; 0, u_2) = x^4/4 + u_2 x^2/2$ $(u_2 < 0)$, whose critical points $x=0$ or $x = \pm\sqrt{u_2}$ corresponding to Morse's maximum (unstable), Morse's positive/negative minimum (stable) respectively. Therefore, $T(Z) = 0 \leq +1 = T(1) \Longrightarrow \Phi(Z) \leqslant_t \Phi(1)$, $T(0) = -1 \leq 0 = T(X) \Longrightarrow \Phi(0) \leqslant_t \Phi(X)$, $T(X) = 0 \leq +1 = T(1) \Longrightarrow \Phi(X) \leqslant_t \Phi(1)$, $T(0) = -1 \leq 0 = T(Z) \Longrightarrow \Phi(0) \leqslant_t \Phi(Z)$.

3) $\Phi$ is involution-preserving: $\Phi(\neg a) = \sigma(\Phi(a))$, where $\sigma(x) = -x$ is the state variable reflection (geometrically, an involutive isometry of the catastrophe unfolding).

Here we should verify the involutive negation $\neg$ in the bilattice on the contravariance with respect to the truth order, i.e. $a \leq_t b \Longrightarrow \neg b \leq_t \neg a$. If $a \leq_t b$ (i.e. $T(a) \leq T(b)$), $T(\sigma(a)) = -T(a) \geq -T(b) = T(\sigma(b))$, then $\sigma(b) \leq_t \sigma(a)$, i.e. $\neg b \leq_t \neg a$.

Consequently, $\Phi$ preserves all algebraic operations derived from these structures. Thus we have established a bilattice isomorphism $\Phi:\mathrm{FOUR} \to \mathrm{Crit}(V)/\mathrm{Morse}(V)$. This mapping not only identifies four logical values with four equivalence classes but also preserves the knowledge order, truth order, and involution, thereby respecting all algebraic operations.

Furthermore, the mapping $\Phi$ can be extended to a categorical equivalence between bilattices and catastrophe unfoldings.

## 4. Bilattice-Catastrophe Isomorphism Theorem

While the functoriality of the twist-product construction is known in lattice theory [6, 7], and the existence of induced maps in stratified spaces is standard [22-24], the explicit correspondence between bilattice homomorphisms and bifurcation-preserving catastrophe morphisms, as detailed in the following theorem, appears to be new.

According to the literature [25], the definition of a bilattice homomorphism directly incorporates the preservation of truth-value structures (i.e. preserving logical order, involution, and logical operation), which is the algebraic basis for constructing induced catastrophe morphisms.

**Definition 4.1** A functor is a categorical equivalence if it is fully faithful and essentially surjective — or equivalently, if it is invertible up to an isomorphism.

**Theorem 4.2** (Induced catastrophe morphism) Given a bilattice homomorphism $h:\mathcal{B}_1 \to \mathcal{B}_2$, there exists a unique catastrophe morphism $\Phi(h): \Phi(\mathcal{B}_1) \to \Phi(\mathcal{B}_2)$ preserving bifurcation set and satisfying the functoriality conditions, where $\Phi(\mathcal{B}_1)=\text{Crit}(V_1)/\text{Morse}(V_1)$, and $\Phi(\mathcal{B}_2)=\text{Crit}(V_2)/\text{Morse}(V_2)$.

***Proof.*** For any $(x,u) \in \text{Crit}(V_1)$, we define $\Phi(h)([x,u])=[y,\phi(u)]$, where $y$ satisfies 1) preserving truth-value structure: $\text{sign}(y)=\text{sign}(h(\text{sign}(x)))$; 2) critical point condition: $(y,\phi(u)) \in \text{Crit}(V_2)$. Thus $\Phi(h)(\text{Crit}(V_1)) \subseteq \text{Crit}(V_2)$, which verifies the preservation of the balanced surface.

Further, if $(x,u) \in \sum_1 \subset \text{Crit}(V_1)$ is a degeneration point, where $\sum_1$ is the bifurcation set of $\mathcal{B}_1$, there are $\frac{\partial V_1}{\partial x}=0$ and $\frac{\partial^2 V_1}{\partial x^2}=0$, which corresponds to $X_1$ (uncertain state) in $\mathcal{B}_1$. Because $h:\mathcal{B}_1 \to \mathcal{B}_2$ preserves truth-value structures, then there is $h(X_1)=X_2$ (the uncertain state in $\mathcal{B}_2$), i.e. $(y, \phi(u)) \in \sum_2$ (the bifurcation set of $\mathcal{B}_2$). Thus $\Phi(h)(\sum_1) \subseteq \sum_2$, which verifies the preservation of the bifurcation set.

Because an identity bilattice homomorphism induces identity parameter mapping and state mapping, $\Phi(\text{id}_\mathcal{B})=\text{id}(\Phi_\mathcal{B})$. Given bilattice homomorphisms $g$ and $h$, according to the functoriality of the construction, the state mapping of a composite is equal to the composite of the state mappings, i.e. $\Phi(g \circ h)=\Phi(g) \circ \Phi(h)$. Thus the functoriality has been verified.

In addition, the equivalence group that preserves the hierarchical structure is a geometric subgroup, thus the basic singularity theory applies, i.e. $\Phi(h)$ also preserves the hierarchical mapping with the same codimensions [26].

**Theorem 4.3** (Bilattice-Catastrophe Isomorphism): The covariant functor $\Phi$ can be further expanded to a categorical equivalence: *Bilat* $\cong$ *Catast* between the category of bilattices and the category of catastrophe unfoldings.

***Proof.*** Any functor that defines an equivalence of categories admits both a left and a right adjoint [27]. For any $\mathcal{B}_1$ and $\mathcal{B}_2$, $\text{Hom}_{\text{Bilat}}(\mathcal{B}_1,\mathcal{B}_2) \cong \text{Hom}_{\text{Catast}}(\Phi(\mathcal{B}_1),\Phi(\mathcal{B}_2))$, because the bilattice homomorphism is determined by its action on the generators, and the generators correspond to the types of critical points of the potential function.

Thus we have established the categorical equivalence *Bilat* $\cong$ *Catast* for the cusp catastrophe (codimension $r=2$), which corresponds to the four-valued bilattice FOUR. For the higher codimension catastrophes ($2 \leq r \leq 4$), the above process can be followed and extended by the same construction, and there exists a categorical equivalence $\Phi_r: Bilat_r \to Catast_r$ between the category of $n$-valued bilattices (where $n=2r$ or $n=2r+2$ depending on symmetry) and the catastrophe category, where $Bilat_r$ is the bilattice category of rank $r$ and $Catast_r$ is the elementary catastrophe category of codimension $r$.

On the whole, the category *Catast* captures catastrophe's geometric-topological structures, while *Bilat* captures algebraic-logical structures. The equivalence *Bilat* $\cong$ *Catast* reveals a deep unity between continuous dynamics and discrete logic.

## 5. Discrete implementation of catastrophe theory

The four-valued logic FOUR=(0, 1, X, Z) can be regarded as a discrete implementation of the catastrophe theory in digital circuits. Any continuous-discrete interface must contain degenerate critical points, because when parameters cross the bifurcation set, the system must go through an "uncertain" state (determined by the delay convention). In addition, when two stable attractors exist simultaneously but the system does not favor either one (local maximum), or the system is completely away from the attractors (high resistance), it corresponds to a logical 'contradiction' or 'floating' state. The absorption law of FOUR corresponds to the structural stability of Morse critical points, i.e. small perturbations do not change the qualitative behavior. Meanwhile, metastability in digital circuits corresponds to degenerate critical points on the bifurcation set, and the unboundedness of its resolution time corresponds to critical slowing down.

Actually the deep connection between FOUR in digital circuits and the catastrophe theory is reflected in: 1) isomorphism of state space. The four values of FOUR correspond precisely to the four qualitative behaviors of the potential function in the catastrophe theory (stable minima, degenerate critical points, unstable points /escape); 2) the topological nature of computation. The absorptive and propagative rules of FOUR are projections of structural stability of the Morse critical points and bifurcation phenomena in catastrophe onto discrete algebraic systems; 3) mathematical unification of uncertainty. The propagation behavior of X (unknown) in circuits has the same algebraic structure as the universal unfolding of degenerate critical points in catastrophe theory, in which both describe the sensitivity and unpredictability of a system at critical parameters; 4) the bridging role of Bilattice. The bilattice structure of Belnap's four-valued logic provides an algebraic framework for connecting discrete many-valued logic with continuous catastrophes, and the Twist-Product construction corresponds to the Splitting Lemma in catastrophe theory.

This connection is not only a mathematical analogy but also reveals the physical reality in digital circuit design: when engineers introduce X and Z states, they are essentially describing phase transitions and critical phenomena in continuous physical systems using the language of discrete algebra, which is precisely the central concern of Thom's catastrophe theory.

We have proved in Theorem 1.2 that FOUR is the minimal complete algebraic structure that describes continuous-discrete interfaces with involution symmetry. Thus engineers introduce X and Z not as 'engineering conveniences', but as an inevitable result of the topological discretization of continuous dynamical systems. Without X, state transitions cannot be described; without Z, disengagement from the drive cannot be described.

Furthermore, by selecting catastrophes with higher codimensions (swallowtail codim=3, butterfly codim=4), five-valued and six-valued logics can be systematically constructed for multi-level storage and quantum computing.

## Conclusion

We have established that the four-valued logic (0, 1, X, Z) fundamental to digital circuit design is the discrete algebraic manifestation of Thom's catastrophe theory.

The correspondence is structural and rigorous: the bilattice organization of truth values mirrors the classification of critical points in potential functions, with X corresponding to degenerate catastrophes and Z to unstable equilibria.

This insight bridges the gap between the continuous physics of electronic devices and the discrete mathematics of digital logic. It provides a theoretical foundation for hardware verification, metastability analysis, and robust design—transforming engineering heuristics into mathematical theorems.

This categorical correspondence provides a foundational framework for unifying the analysis of nonlinear system dynamics, uncertain/contradictory information processing, and multi-valued reasoning.

Future work will explore stochastic catastrophe theory for noise analysis, quantum catastrophe theory for quantum computing architectures, and algorithmic catastrophe detection for automated verification tools.